# Rapid and sensitive detection of SARS-CoV-2 with functionalized magnetic nanoparticles

Jing Zhong*, Enja Laureen Rösch, Thilo Viereck, Meinhard Schilling and Frank Ludwig

Institute for Electrical Measurement Science and Fundamental Electrical Engineering and Laboratory for Emerging Nanometrology (LENA), TU Braunschweig, Hans-Sommer-Str. 66, Braunschweig D-38106, Germany

*Corresponding author: j.zhong@tu-braunschweig.de

**Abstract**

The outbreak of the severe acute respiratory syndrome coronavirus 2 (SARS-CoV-2) threatens global medical systems and economies, and rules our daily living life. Controlling the outbreak of SARS-CoV-2 has become one of the most important and urgent strategies throughout the whole world. As of October, 2020, there have not yet been any medicines or therapies to be effective against SARS-CoV-2. Thus, rapid and sensitive diagnostics is the most important measures to control the outbreak of SARS-CoV-2. Homogeneous biosensing based on magnetic nanoparticles (MNPs) is one of the most promising approaches for rapid and highly sensitive detection of biomolecules. This paper proposes an approach for rapid and sensitive detection of SARS-CoV-2 with functionalized MNPs via the measurement of their magnetic response in an ac magnetic field. Experimental results demonstrate that the proposed approach allows the rapid detection of mimic SARS-CoV-2 with a limit of detection of 0.084 nM (5.9 fmole). The proposed approach has great potential for designing a low-cost and point-of-care device for rapid and sensitive diagnostics of SARS-CoV-2.

**Keywords:** SARS-CoV-2, magnetic nanoparticles, magnetic particle spectroscopy, limit of detection

# 1. Introduction

Severe acute respiratory syndrome coronavirus 2 (SARS-CoV-2) is an RNA based virus that causes coronavirus disease 2019 (COVID-19) [1]. SARS-CoV-2 with a dimeter ranging from about 60 to 140 nm has spike, envelope, membrane and nucleocapsid proteins [2, 3]. Spike protein is the one for allowing the virus to bind to the receptor ACE2 of a host cell and to enter the host cell [4-8]. A patient infected with SARS-CoV-2 may suffer from severe pneumonia and acute respiratory distress syndrome [3, 9, 10]. Since the first infection case of SARS-CoV-2 reported in December, 2019 [10, 11], SARS-CoV-2 has spread across the whole world, which causes severe problems in medical systems, economics and other social issues. Till early October 2020, World Health Organization (WHO) has reported more than 35 million cases of infection and 1 million deaths in the world [12]. In addition to other control measures, e.g. social distance and hand washing, WHO has strongly recommended a large amount of testing not only of symptomatic persons and their contacts but also of asymptomatic contacts. Therefore, large amounts of tests are one of the most important measures to control the outbreak of SARS-CoV-2.

To date, polymerase chain reaction (PCR) based tests with high sensitivity and specificity are the gold standard for the diagnostics of SARS-CoV-2 infection [13, 14]. PCR tests include several complex experimental procedures, which costs at least a few hours to get the test results. In a PCR test, most experimental procedures require sophisticated and very expensive instrumentation and should be handled by experienced experts. Otherwise, there is a high chance for a false positive/negative result. It leads to huge efforts for medical staffs in the clinics, especially during the exponential increase in infection cases. In developing and underdeveloped countries, the situation for the diagnostics of SARS-CoV-2 with PCR is even worse due to the lack of equipped labs, instrumentation and experienced experts. Therefore, it is of extreme importance and necessity to develop new methods and/or instrumentation for rapid diagnostics of SARS-CoV-2 with reasonable costs. WHO highly encourages researches on the performance of rapid diagnostics approaches and potential diagnostic utilities. However, the development of new approaches for rapid diagnostics, in addition to PCR-based tests, is still a challenging research topic.

Homogeneous biosensing based on magnetic nanoparticles (MNPs) is one of the most promising approaches for rapid and sensitive detection of specific biomolecules, e.g. protein, DNA/RNA and virus. It employs the change in the magnetization dynamics and consequent magnetic response of the MNPs exposed to time-varying magnetic fields [15]. The binding behavior of the biomolecules to functionalized MNPs, dominated by Brownian relaxation, increases their hydrodynamic size or forms cross-linking structures, which significantly changes the Brownian relaxation time of the MNPs and their dynamic magnetization in time-varying magnetic fields [16-18]. For instance, for MNPs bound with biomolecules, the harmonics in magnetic particle spectroscopy (MPS) dominated by Brownian relaxation decrease when exposed to a sufficiently strong ac magnetic field. Especially, higher harmonics decrease faster than the fundamental harmonic, thus showing a decrease in the harmonic ratio that is independent of the MNP concentration [16, 19]. Thus, the measurement of the MNP magnetization and its susceptibility/spectra in ac magnetic fields can be used to detect the quantity of specific biomolecules. An MPS system has been demonstrated to be a low-cost, versatile and sensitive tool to measure MNP magnetization and dynamics for biomolecule detection [16, 20]. For instance, a mixing-frequency MPS system was designed to measure the mixing components of protein G-functionalized MNP magnetization for antibody detection [21]. Zhang *et al.* reported on the measurement of the harmonic ratio of anti-thrombin DNA aptamers-functionalized MNPs for the detection of thrombin [22]. Zhong *et al.* investigated the effect of binding behavior on MNP relaxation time and harmonic ratio for the detection and imaging of biotinylated IgG using streptavidin functionalized MNPs [16, 19]. Yang et al. demonstrated the feasibility of wash-free, sensitive and specific assays for the detection of different viruses, e.g. orchid and influenza viruses, with antibody-functionalized MNPs by measuring the reduction in the ac susceptibility in mixed-frequency ac magnetic fields [23-25]. Tian *et al.* reported on homogeneous detection of SARS-CoV-2 utilizing an opto-magnetic measurement system for the detection of the RND-dependent RNA polymerase coding sequence of SARS-CoV-2 [26]. All these approaches have demonstrated that MNP-based homogeneous biosensing is a wash-free and mix-and-measure approach for rapid and sensitive detection of specific biomolecules. Therefore, it is of great interest and importance to investigate MNP-based biosensing for rapid and sensitive detection of SARS-CoV-2.

In this paper, we propose the detection of SARS-CoV-2 via the measurement of the MPS signal of SARS-CoV-2 spike protein antibody-functionalized MNPs. For the proof-of-concept, SARS-CoV-2 spike protein antibody-functionalized MNPs are used as sensors to detect mimic SARS-CoV-2 that is 100 nm polystyrene bead conjugated with SARS-CoV-2 spike protein. A custom-built MPS system is used to measure the MPS signal of the functionalized MNPs. In addition, the ac susceptibility (ACS) spectra are measured with a rotating magnetic field (RMF) system for the evaluation of the Brownian relaxation time. Experiments on samples with different mimic virus concentrations are performed to evaluate the measurement sensitivity and limit of detection (LOD) of the present approach.

## 2. Materials and Methods

In this paper, functionalized MNPs dominated by Brownian relaxation are used to detect a mimic SARS-CoV-2. The functionalized MNPs consist of BNF-80 nanoparticles coated with protein A, purchased from micromod Partikeltechnologie GmbH (Rostock, Germany), and SARS-CoV-2 spike protein monoclonal antibodies (anti-SARS-CoV-2 spike), purchased from Biomol GmbH (Hamburg, Germany), as show in Fig. 1a. Due to the high binding affinity between protein A and the antibody Fc fragment, SARS-CoV-2 spike protein antibodies are easily conjugated onto the surface of protein A-coated BNF-80 nanoparticles. BNF-80 nanoparticles with protein A have an original iron concentration of 5.5 mg/ml and molar particle concentration of 20 pmole/mL (20 nM). The stock BNF-80 nanoparticle sample is diluted to 2 nM for experiments. The mimic SARS-CoV-2 consists of streptavidin-coated polystyrene beads (purchased from micromod Partikeltechnologie GmbH) and biotinylated SARS-CoV-2 spike proteins (RBD-SD1, Avi-His Tag, from Biomol GmbH), as show in Fig. 1b. Due to the high binding affinity between streptavidin and biotin, the biotinylated SARS-CoV-2 spike proteins are conjugated onto the surface of streptavidin-functionalized polystyrene beads. The polystyrene beads have a nominal size of 100 nm, which is similar to the size of SARS-CoV-2 [27].

The functionalized MNPs and mimic SARS-CoV-2 are prepared and incubated for more than 12 hours prior to experiments. To make sure the conjugation between the antigen and antibody is finished, experiments on the measurements of ACS spectra and MPS signal are performed more than 1 hour after mixing the functionalized MNPs and the mimic SARS-CoV-2. Totally, five experimental samples with different mimic virus concentrations from 0 to 3.38 nM are prepared. Note that in this paper all the mole quantity/concentration of mimic virus is given by that of polystyrene beads. Table I lists all the details regarding the amounts of BNF-80 nanoparticles, antibody, spike protein and polystyrene beads for sample preparation. Each sample contains the same-concentration (2 nM) of functionalized MNPs. For the functionalized MNPs, the quantity of the SARS-CoV-2 spike protein antibody is 4 pmole, which is – according to preliminary tests – a balance between the binding affinity of the functionalized MNPs and an increase in their hydrodynamic sizes. For control experiments, the protein A-coated BNF-80 nanoparticles without SARS-CoV-2 spike protein antibody are used to study the non-specific binding between the unfunctionalized MNPs and the mimic SARS-CoV-2 as well as an increase in viscosity due to the presence of the mimic SARS-CoV-2. The samples for control experiments are prepared in the same way as that for measurements, except the absence of the SARS-CoV-2 spike protein antibodies.

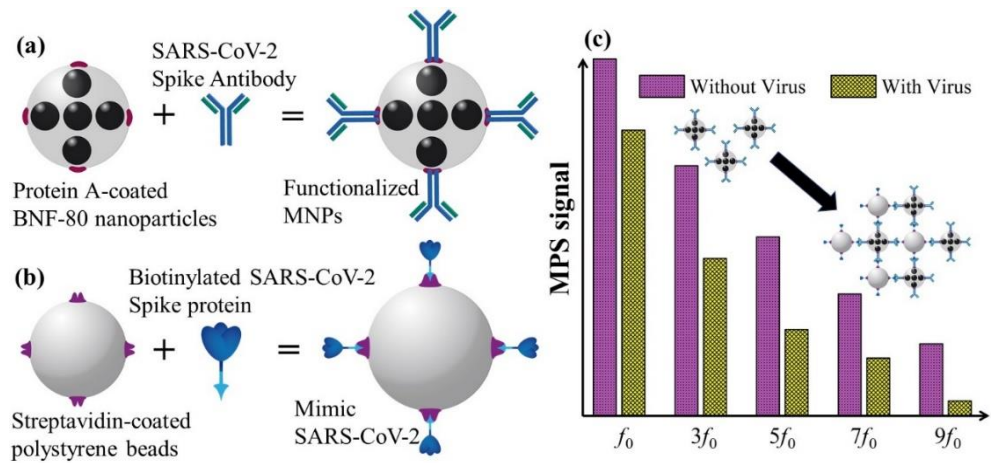

Fig. 1. (a) Schematic of functionalized MNPs, (b) schematic of mimic SARS-CoV-2, (c) Schematic of MPS signal with and without mimic virus.

Table I. Sample preparation for rapid detection of mimic SARS-CoV-2.

| | Functionalized MNPs | | Mimic SARS-CoV-2 | | Total volume | SARS-CoV-2 concentration |
|---|---|---|---|---|---|---|
| | BNF-80 with protein A | SARS-CoV-2 spike protein antibody | Polystyrene beads | Biotinylated spike protein | | |
| S1 | 0.18 pmole | 4 pmole | 0.000 pmole | 0.0 pmole | 90 $\mu$l | 0.00 nM |
| S2 | 0.18 pmole | 4 pmole | 0.076 pmole | 7.6 pmole | 90 $\mu$l | 0.84 nM |
| S3 | 0.18 pmole | 4 pmole | 0.152 pmole | 15.2 pmole | 90 $\mu$l | 1.68 nM |
| S4 | 0.18 pmole | 4 pmole | 0.228 pmole | 22.8 pmole | 90 $\mu$l | 2.53 nM |
| S5 | 0.18 pmole | 4 pmole | 0.304 pmole | 30.4 pmole | 90 $\mu$l | 3.38 nM |

Exposed to a sufficiently strong ac magnetic field with excitation frequency $f_0$, the MPS signal consists of not only the fundamental harmonic at frequency $f_0$ but also higher harmonics at frequency $i \times f_0$. Fig. 1c shows the schematic of the MPS signal of the MNPs dominated by Brownian relaxation. In absence of mimic SARS-CoV-2, the functionalized MNPs can freely rotate, thus resulting in rich harmonic spectra. With the presence of mimic SARS-CoV-2, the functionalized MNPs bind with the mimic SARS-CoV-2 due to the specific binding behavior between antibody and antigen. This leads to an increase in effective hydrodynamic size of the MNPs. Consequently, the functionalized MNPs cannot freely rotate or rotate slowly in response to the excitation magnetic field, which causes a significant decrease in the MPS signal. Therefore, the quantity of mimic SARS-CoV-2 can be measured and quantified by the measurement of MPS signal of the functionalized MNPs. In this paper, a detection coil-based scanning MPS system is used to measure the MPS signal of the functionalized MNPs for rapid detection of mimic SARS-CoV-2 in 10 mT-excitation magnetic field in the frequency range from 140 Hz to 1.4 kHz [28]. The measurement time for MPS amounts to about 9 seconds, including 1 second for sample measurement, 1 second for blank measurement and 7 seconds for mechanical movement. In addition, a fluxgate sensor-based RMF system is used to measure the ACS spectra in a 0.5 mT-excitation magnetic field in the frequency range from 1 Hz to 3 kHz [29]. The sample volumes for the measurements of ACS spectra and MPS are 90 $\mu$L and 70 $\mu$L, respectively.

# 3. Results

## 3.1 AC Susceptibility

Figure 2 shows the measured ACS spectra of functionalized and unfunctionalized MNPs with different mimic virus concentrations. Fig. 2a shows the measured imaginary parts of the ACS spectra on the functionalized MNPs. With increasing mimic virus concentration, the shape of the ACS spectra gets broader and the peak frequency shifts to a lower frequency. This indicates an increase in the effective Brownian relaxation time $\tau_B$ [16]. Fig. 2b shows the measured imaginary parts of the ACS spectra for the unfunctionalized MNPs. The shape of the imaginary parts shown in Fig. 2b slightly shifts to lower frequencies with increasing mimic virus concentration. It may be caused by non-specific binding between the mimic virus and the unfunctionalized MNPs, as well as an increase in the solution viscosity due to the presence of the mimic virus. The changes in the ACS spectra of the functionalized MNPs are much more significant than those of the unfunctionalized MNPs, indicating that the specific binding behaviors between the SARS-CoV-2 antigen and antibody dominate.

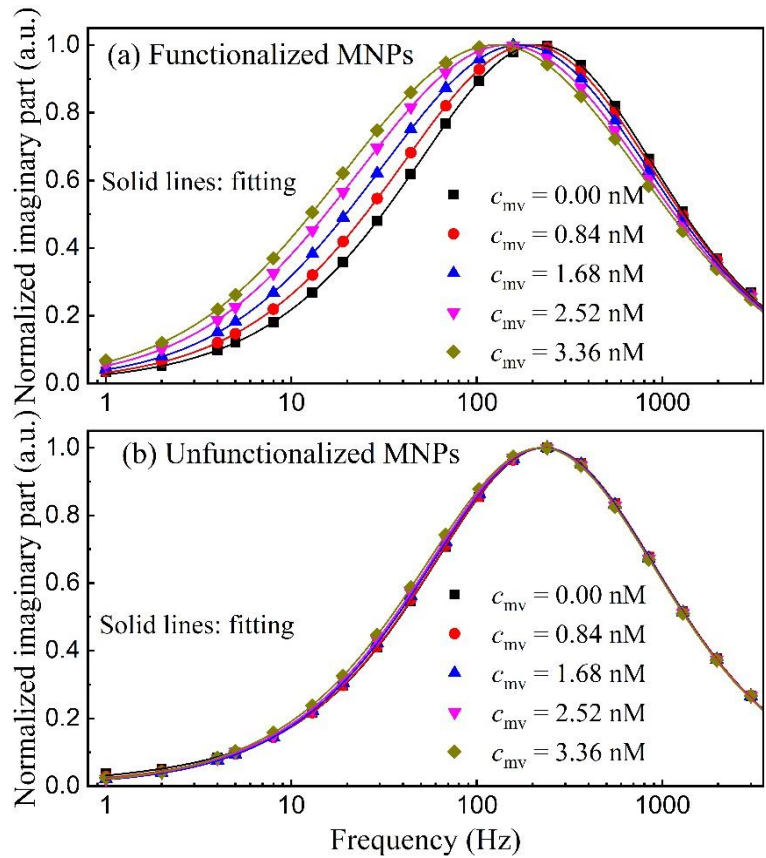

Fig. 2. Experimental results of ac susceptibility spectra of functionalized (a) and unfunctionalized (b) MNPs with different mimic virus concentration. Symbols are experimental data whereas solid lines are fitting with a generalized Debye model with bi-lognormal size distribution.

To quantitatively investigate the dependence of the effective Brownian relaxation time $\tau_B$ on the mimic virus concentration, $\tau_B$ was estimated from the peak frequency of the ac susceptibility imaginary part. To precisely determine the peak frequency despite the limited number of data points, the imaginary parts of the ACS spectra were fitted with a generalized Debye model with a bimodal lognormal size distribution. The solid lines in Fig. 2 shows that the generalized Debye model fits the experimental data very well. Then, $\tau_B$ was calculated via the peak frequency $f_{peak}$ of the maximum in the fitted imaginary parts applying $\tau_B = 1/(2 \times \pi \times f_{peak})$. Fig. 3 shows the dependence of the effective Brownian relaxation time $\tau_B$ on the mimic virus concentration of functionalized and unfunctionalized MNPs. For the unfunctionalized MNPs, $\tau_B$ increases by a factor of about 1.05 (from 0.687 to 0.719 ms) whereas for functionalized MNPs $\tau_B$ significantly increases by a factor of about 1.74 (from 0.742 to 1.292 ms). Additional control experiments on ACS spectra were also performed with BNF-80 nanoparticles functionalized with some other antibodies (not specific to SARS-CoV-2 spike protein) and the mimic virus (data not shown), which also show minor changes in the effective Brownian relaxation time. This indicates that the specific binding between SARS-CoV-2 antigen and antibody significantly increases the effective Brownian relaxation time $\tau_B$. Note that, for a mimic virus concentration $c_{mv} = 0$ nM, $\tau_B$ for the functionalized MNPs is greater than that for the unfunctionalized MNPs, which is caused by a larger hydrodynamic size of the functionalized MNPs due to the conjugation of SARS-CoV-2 spike protein antibodies onto the surface of the protein A-coated BNF-80 nanoparticles. Therefore, our experimental results demonstrate that the specific binding of the mimic SARS-CoV-2 onto the surface of the functionalized MNPs significantly increases the effective Brownian relaxation time.

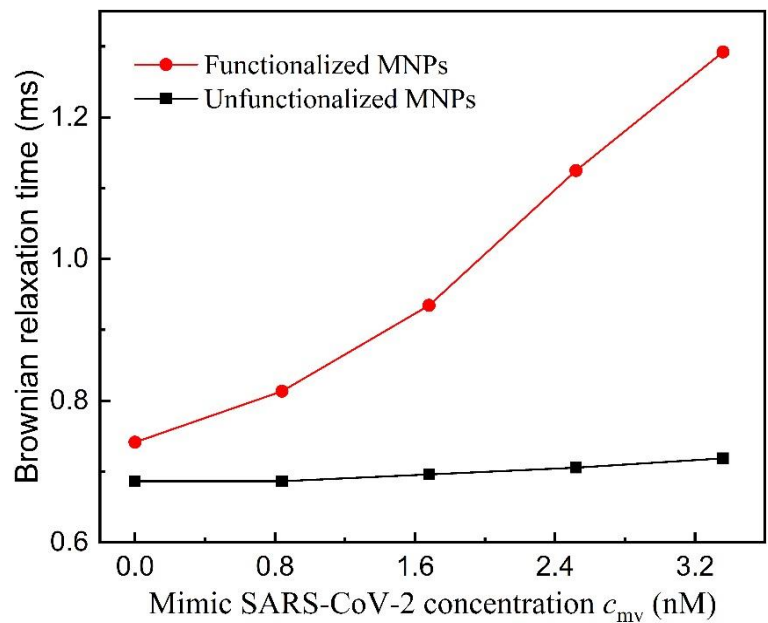

Fig. 3. Experimental results of effective Brownian relaxation time vs. mimic SARS-CoV-2 concentration. Symbols are experimental data whereas solid lines are guides to the eyes.

## 3.2 Magnetic Particle Spectroscopy

The custom-built SMPS was used to measure the 1$^{st}$ and 3$^{rd}$ harmonics ($M_1$ and $M_3$) of the output voltage from the gradiometric detection coils on experimental samples in ac magnetic fields with different excitation frequencies $f_0$, where the harmonic at frequency $i \times f_0$ is defined as $i^{th}$ harmonic $M_i$. Figure 4 shows the experimental results of the harmonic ratio $R_{3rd/1st} = M_3/M_1$ vs. frequency curves on samples with different mimic virus concentrations $c_{mv}$. Note that the harmonic ratio is independent of the MNP concentration, but only dependent on the mimic virus concentration $c_{mv}$ in a given-frequency ac magnetic field. Figures 4a and 4b show the experimental results of functionalized and unfunctionalized MNPs, respectively. Without SARS-CoV-2 spike protein antibody functionalization, the harmonic ratio $R_{3rd/1st}$ of the MNPs coated with protein A slightly decreases with increasing mimic virus concentration $c_{mv}$, which might be caused by non-specific binding and viscosity changes due to the presence of the mimic virus. On the other hand, with SARS-CoV-2 spike protein antibody functionalization, the harmonic ratio $R_{3rd/1st}$ of the functionalized MNPs decreases significantly with an increase in the mimic virus concentration $c_{mv}$. Moreover, the changes in the $R_{3rd/1st}$ decrease with increasing excitation frequency, which agrees well with published results [16]. By comparing Figs. 4a and 4b, the amounts of mimic SARS-CoV-2 can be quantified with the harmonic ratio $R_{3rd/1st}$ at each single excitation frequency.

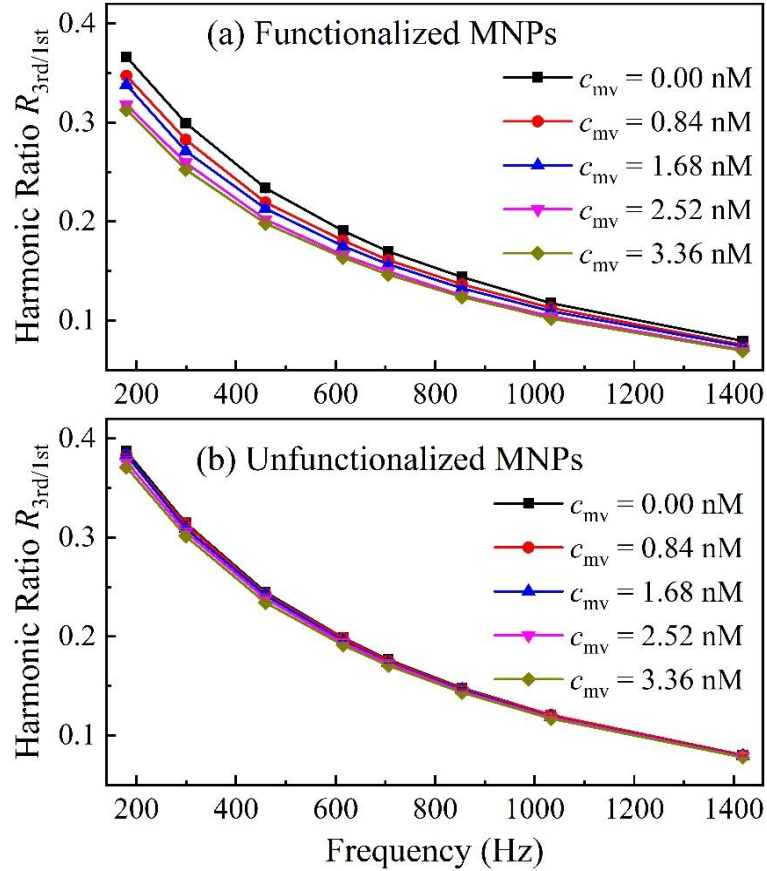

Fig. 4. Experimental results of measured harmonic ratio $R_{3rd/1st}$ vs. excitation frequency of functionalized (a) and unfunctionalized (b) MNPs with different mimic virus concentration $c_{mv}$. Experimental results are averaged for 4 independent repeat measurements. Symbols are experimental results whereas solid lines are guides to the eyes.

To quantitatively evaluate the changes in $R_{3rd/1st}$ vs. mimic SARS-CoV-2 concentration $c_{mv}$, the absolute change $\Delta R_{3rd/1st} = R_{3rd/1st}(c_{mv} = 0 \text{ nM}) - R_{3rd/1st}(c_{mv})$ was calculated and presented in Fig. 5. Figure 5a and 5b show $\Delta R_{3rd/1st}$ vs. mimic SARS-CoV-2 concentration $c_{mv}$ curves of functionalized and unfunctionalized MNPs at different excitation frequencies, respectively. It clearly shows that the changes in $\Delta R_{3rd/1st}$ vs. $c_{mv}$ on the functionalized MNPs are much more significant than those on the unfunctionalized MNPs. With decreasing mimic virus concentration $c_{mv}$, the $\Delta R_{3rd/1st}$ values of the unfunctionalized MNPs decrease as well, indicating that the non-specific binding or change in viscosity becomes less relevant at a low mimic virus concentration. The percentage of $\Delta R_{3rd/1st}$ for unfunctionalized MNPs to that for functionalized MNPs is less than 30%. In addition, it decreases to less than 20% with $c_{mv}$ =

0.84 nM. Additional control experiments on the MPS signal were also performed with BNF-80 nanoparticles functionalized with some other antibodies (not specific to SARS-CoV-2 spike protein) and the mimic virus (data not shown), which also show minor changes in $\Delta R_{3rd/1st}$. It means that the change in $\Delta R_{3rd/1st}$ is dominated by the specific binding between the antibody and antigen. Our experimental results demonstrate that the harmonic ratio $R_{3rd/1st}$ can be used for the detection of mimic SARS-CoV-2.

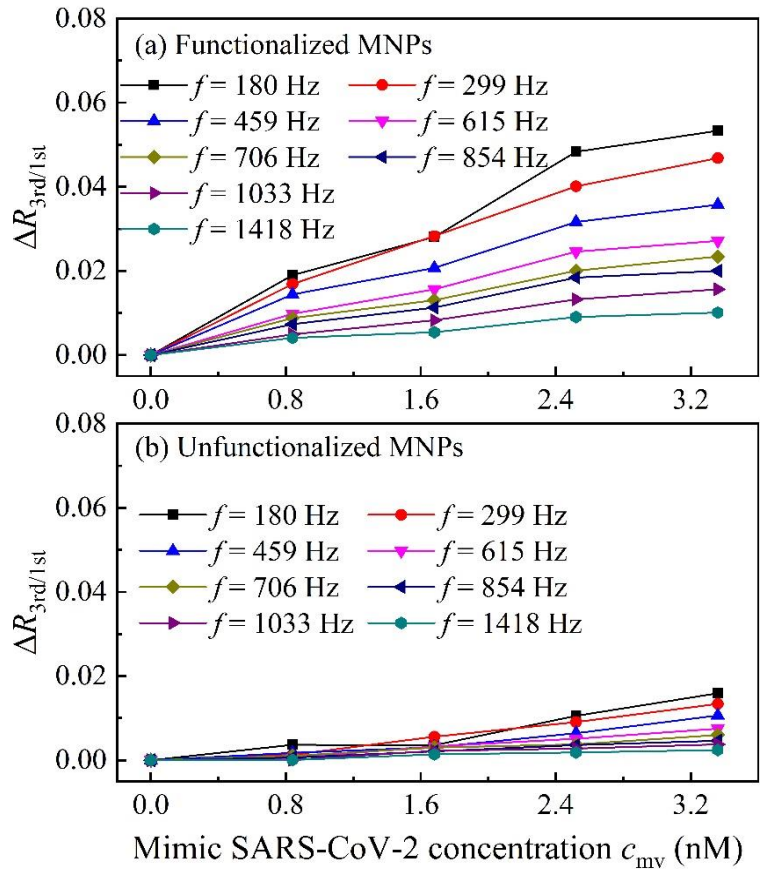

Fig. 5. Absolute variation in measured harmonic ratio $\Delta R_{3rd/1st}$ vs. mimic SARS-CoV-2 concentration of functionalized (a) and unfunctionalized (b) MNPs at different excitation frequencies. Symbols are experimental results whereas solid lines are guides to the eyes.

### 3.3 Limit of Detection

The limit of detection (LOD) is one of the most important measures for sensitive detection. To evaluate the LOD, the measurement sensitivity $\eta$ and noise level on the harmonic ratio are calculated. The measurement sensitivity, defined as the relative change in $\Delta R_{3rd/1st}$ to $c_{mv}$ $d\Delta R_{3rd/1st}/dc_{mv}$, is calculated by a linear fitting of the $\Delta R_{3rd/1st}$ vs. $c_{mv}$ curves shown in Fig. 5. Figure 6a shows the measurement sensitivity vs. frequency curve for different excitation frequencies. It indicates that with increasing frequency from 140 to 1418 Hz the measurement sensitivity $\eta$ decreases from 0.017 nM$^{-1}$ to 0.004 nM$^{-1}$, which quantitatively fits well with published results [16]. The noise level is calculated with the standard deviation $\varepsilon$ of 10-repeated measurements on the harmonic ratio of an MNP sample with $c_{mv}$ = 0 nM. In principle, with increasing the excitation frequency, the signal-to-noise ratio (SNR) of the measured harmonic gets improved, which would improve the standard deviation in measured harmonic ratio due to Faraday's law. However, the data in frequencies of 299 Hz and 459 Hz show higher noise levels than that in 180 Hz, which may be caused by instabilities of the measurement system. Figure 6b shows the LOD $\delta$, estimated from the measurement sensitivity $\eta$ and standard deviation $\varepsilon$ by $\delta = 3.3\varepsilon/\eta$ [30, 31]. It indicates that the LOD is in the range from about 0.10 to 0.37 nM. Note that the estimated LOD is obtained for a single measurement. With $n$-time repeated measurements, the LOD can be further improved by a factor of $\sqrt{n}$.

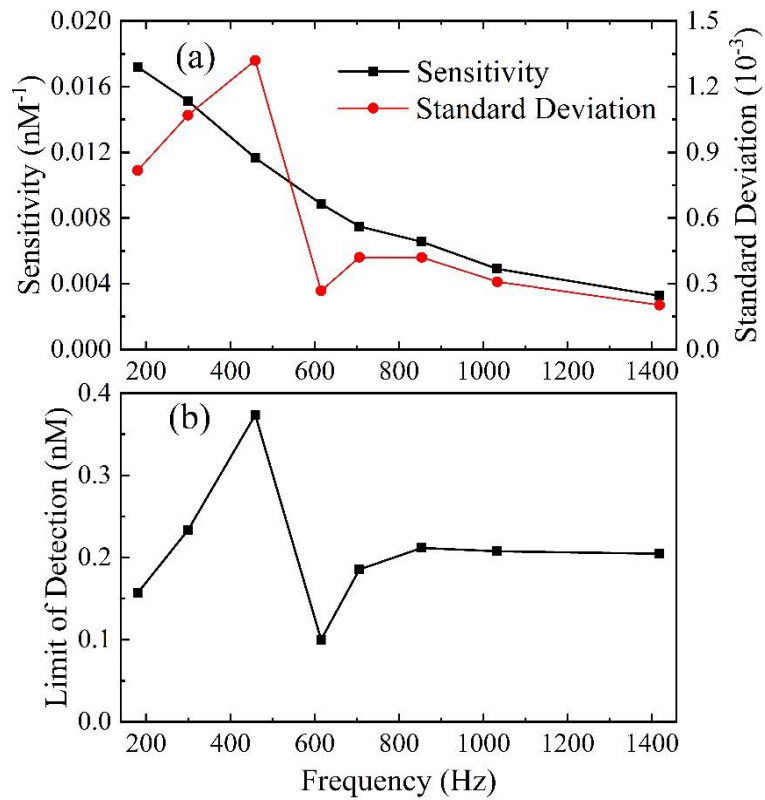

Fig. 6. (a) shows the measurement sensitivity of mimic SARS-CoV-2 (left axis) and standard deviation (right axis) vs. frequency curves. (b) shows the calculated limit of detection.

To experimentally investigate the LOD of mimic SARS-CoV-2 with the present approach, several samples with low mimic virus concentrations are prepared. Figure 7 shows the $\Delta R_{3rd/1st}$ vs. $c_{mv}$ curves measured at different frequencies. The experimental data were averaged from 4 repeated measurements with a total measurement time of about 36 seconds (note that each of the 4 cycles includes 1 s measurement on the MNP sample, 7 s mechanical movement of the robot and 1 s background measurement). It clearly shows that for all the frequencies the lowest mimic virus concentration of about 0.084 nM is detectable, which is slightly better than the estimated LOD from 0.05 to 0.18 nM with 4 repeated measurements. It might be caused by the underestimation of the measurement sensitivity since the measurement sensitivity is not obtained at very low concentration of mimic virus. The present approach allows an LOD in terms of mimic virus concentration of about 0.084 nM with a sample volume of about 70 $\mu$l and an LOD in terms of mole quantity of about 5.9 fmole.

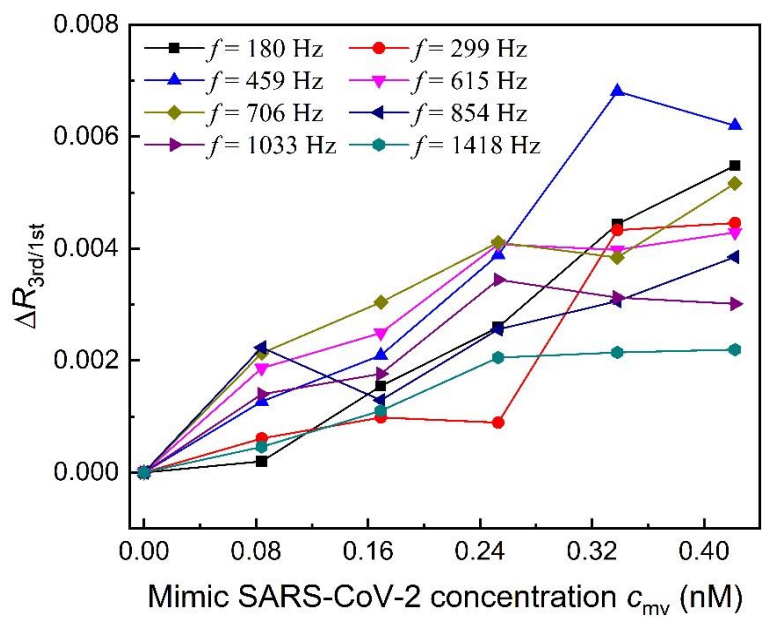

Fig. 7. Experimental results of $\Delta R_{3rd/1st}$ vs. mimic SARS-CoV-2 concentration at different frequencies. Experimental results are the average of 4-repeated measurements. Symbols are experimental data whereas solid lines are guides to the eye.

## 4. Discussion

Rapid and highly sensitive detection of SARS-CoV-2 is of great importance to control the outbreak of SARS-CoV-2, thus improving the pandemic situation. This paper proposed the detection of SARS-CoV-2 with functionalized MNPs via the measurements of MPS signal, specifically the ratio of the 3$^{rd}$ to 1$^{st}$ harmonics. For a proof-of-concept, we used 100 nm-size polystyrene beads conjugated with SARS-CoV-2 spike protein to mimic SARS-CoV-2 while SARS-CoV-2 spike protein-functionalized MNPs are used as sensors. Our experimental results show that the present approach allows a highly sensitive detection of mimic SARS-CoV-2 with a LOD of about 0.084 nM (5.9 fmole) with 4-repeated measurements in about 36 seconds, including 4 seconds for measurement on an MNP sample, 28 seconds for mechanical movement and 4 seconds for background measurement. Note that the MPS system was built for two-dimensional imaging. With a tailored measurement system, we believe that the LOD can be further improved without any significant costs. For instance, a tailored MPS system can measure 36 averages of the MPS signal within a total measurement time of 36 s if a blank measurement is performed in advance. In this case, the LOD can be improved by a factor of $\sqrt{n}$ with $n$ repeated measurements.

WHO has developed the ASSURED criteria as a benchmark to evaluate if tests address disease control needs: **A**ffordable, **S**ensitive, **S**pecific, **U**ser-friendly (simple to perform in a few steps with minimal training), **R**apid and robust (results available in less than 30 min), **E**quipment-free and **D**eliverable to end-users [32, 33]. However, rare test methods can fit all the criteria in practice. MNP-based homogenous biosensing only requires the mixture of the functionalized MNPs and the sample to be test without any additional washing steps. The test results can be available in some seconds without taking into account the time required for antigen and antibody conjugation. Finally, the consuming time for the test results with the approach of MNP-based homogenous biosensing mainly depends on the kinetic of SARS-CoV-2 antigen and antibody, which is in the order of about 10 min. It means that the present approach is sensitive, specific, and rapid and robust. The required materials – functionalized MNPs – are affordable, harmless and deliverable to end-users. Thus, it fits all the ASSURED criteria except Equipment-free. However, the measurement setup – an MPS system, which consists of Helmholtz/solenoid coils for the generation of magnetic fields, pick-up coils for the detection of MNP magnetization, some basic electronics for power amplifiers, pre-amplifiers and analog-to-digital/digital-to-analog converters, can be built as a point-of-care (POC) device with a low cost. For ultra-sensitive detection of the analyte with the absolute signal from the magnetic response of functionalized MNPs, e.g. susceptibility and harmonic amplitude, the LOD may be dominated by the preparation error in the MNP concentration for measurements. The current approach employs the harmonic ratio of the $3^{rd}$ to the $1^{st}$ harmonics for bio-sensing, which is independent of MNP concentration but only dependent on the analyte concentration, thus eliminating the preparation error in the MNP concentration. Therefore, we believe that the present approach is one of the most promising approaches for rapid diagnostic of SARS-CoV-2 in terms of the ASSURED criteria.

For SARS-CoV-2 – this new virus, there have been still a lot of open questions to be investigated. For instance, it is still an open question that how many living viruses there are in a saliva sample and how the infectious possibility is for asymptomatic infection cases. Current approaches for RNA detection only give the information on the RNA concentration in a saliva sample. It means that they can only distinguish that some patients have been infected or not, but not the infection status. For instance, RNA-based test may not distinguish some infected patients with self-recovery who may just have some left virus RNA but no living viruses. An approach of living virus detection will give some further information on the quantity of living viruses in a saliva sample, which is of great importance for medical doctors to evaluate the infectiousness. We believe that the present approach allows highly sensitive, rapid and quantitative detection of living virus, which allows the evaluation of infection status. Especially, combination of RNA and living virus detection will significantly contribute to the understanding of the infection status and infectiousness. In addition, for the functionalized MNPs bound with 100 nm viruses, the change in the hydrodynamic size is much more significant than that for the functionalized MNPs bound with spike proteins. Thus, the direct detection of the virus, in principle, is more sensitive than that of the spike protein. Furthermore, the current approach employs the measurement of magnetic signal of the functionalized MNPs, which does not have depth limitation. When combining with multi-color magnetic particle imaging [33-35], the present approach can be extended to visualize the spatial distribution of SARS-CoV-2 in vivo, which is of great significance and interest not only to control the outbreak of SARS-CoV-2 but also to fundamental researches, e.g. understanding the underlying mechanisms of the infection process and virus proliferation.

## 5. Conclusion

This paper investigated rapid and sensitive detection of SARS-CoV-2 with functionalized MNPs. For a proof-of-concept, functionalized MNPs were used as sensors to detect a mimic virus consisting of 100 nm-polystyrene beads conjugated with SARS-CoV-2 spike proteins. Experiments on ACS spectra and MPS signal of samples with different mimic virus concentrations were performed. Experimental results showed that the binding behavior between mimic SARS-CoV-2 and functionalized MNPs increases the effective Brownian relaxation time and changing the MPS signal. The change in the ratio of the $3^{rd}$ to $1^{st}$ harmonics with the mimic virus concentration was used to analyze the measurement sensitivity and limit of detection. We believe that the proposed approach is of great promise to highly sensitive and rapid detection with a low cost, easy handling of the sample to be detected (mix-and-measure). We envisage that the present work is of great interest and significance to develop new methods and design point-of-care devices for rapid diagnostics of SARS-CoV-2 to control its outbreak, as well as fundamental researches on the virus infection.

## Acknowledgements

Financial supports from the German Research Foundation DFG (Project no.: ZH 782/1-1) and the DFG Research Training Group 1952 Metrology for Complex Nanosystems are gratefully acknowledged. Furthermore, we thank Dr. Esther Wezel, Dr. Giulio Russo and Dr. Wei He from TU Braunschweig, and Dr. Juhao Yang from Helmholtz Centre for Infection Research for very interesting and helpful discussions.

**Conflicts of Interest**

The authors declare no conflicts of interest.